# Single-shot multispectral imaging with a monochromatic camera

SUJIT KUMAR SAHOO,[1,2,*] DONGLIANG TANG,[1] AND CUONG DANG[2,*]

[1]*Centre for OptoElectronics and Biophotonics (OPTIMUS), School of Electrical and Electronic Engineering, The Photonic Institute (TPI), Nanyang Technological University Singapore, 50 Nanyang Avenue, 639798, Singapore*
[2] *Department of Statistics and Applied Probability, National University of Singapore, Singapore*
*Corresponding author: sujit@pmail.ntu.edu.sg, hcdang@ntu.edu.sg*

**Multispectral imaging plays an important role in many applications from astronomical imaging, earth observation to biomedical imaging. However, the current technologies are complex with multiple alignment-sensitive components, predetermined spatial and spectral parameters by manufactures. Here, we demonstrate a single-shot multispectral imaging technique that gives flexibility to end-users with a very simple optical setup, thank to spatial correlation and spectral decorrelation of speckle patterns. These seemingly random speckle patterns are point spreading functions (PSFs) generated by light from point sources propagating through a strongly scattering medium. The spatial correlation of PSFs allows image recovery with deconvolution techniques, while the spectral decorrelation allows them to play the role of tune-able spectral filters in the deconvolution process. Our demonstrations utilizing optical physics of strongly scattering media and computational imaging present the most cost-effective approach for multispectral imaging with great advantages.**

Multispectral imaging has been developed rapidly [1,2] and become an important technology for various applications [3-5] because in addition to the spatial domain, the spectral domain contains significant amount of information about objects [6,7]. One can do it simply by taking multiple shots (time multiplexing) with multiple filters in front of a monochromatic camera [8]. With development of high-resolution cameras, it is practical to trade off spatial resolution to gain spectral information in a single shot imaging technique (space multiplexing). One version of multispectral imaging devices is the colour camera where the multiple spectral filters are spatially distributed on a 2D detector array. One can even rid of the spatial resolution to achieve higher spectral resolution with a large number of spectral filters to use as a compact spectrometer [9]. Fabricating these different spectral filters is difficult and therefore high spectral resolution is challenging.

Utilizing the optical diffractive or refractive components with computational techniques can achieve higher spectral resolution [10-12]. However, the trade-off between spectral resolution and spectral range can limit their performance, especially when the spectrum is just a few narrow lines in a broad range. These technologies all require special optical devices and critical alignment with a camera in a complex optical system where the spatial and spectral information trade-off as well as spectral resolution and spectral range trade-off are pre-set at manufacture time [13,14]. In this article, we demonstrate the simplest method that utilizes a strongly scattering medium to retrieve both spatial and spectral information from a single 2D image. The trade-off between spatial and spectral information including spectral resolution and spectral range is up to analyser's choices to utilize the full capacity of 2D imager.

Light passing through a scattering medium produces a random speckle pattern [15], which has seemingly no spatial and spectral information of the underlying object [Fig. 1(a) and 2(b)]. For imaging through scattering media, a number of techniques have been introduced to reduce or eliminate the scattering effect [16,17] such as adaptive optics [18], wave-front shaping [19], multiphoton fluorescent imaging [20] or optical coherence tomography (OCT) [21]. Recently, transfer matrix inversion method, which maps inputs to outputs of a scattering medium by a series of measurements, utilizes the scattering medium as a scattering lens for imaging [22,23]. The ability of a scattering lens in capturing light with large transverse momentum allows very high-resolution imaging [24]. There are too many propagation modes for light transmitting through a scattering medium. Therefore, good characterization of these modes for just a single wavelength in the mapping process is time consuming with huge data and then heavy computation is required for the image reconstruction process.

However, the transmitted light through a scattering medium is not completely random, in fact, it "remembers" the original direction in a limited range [15,25,26]. This memory effect implies that when the light incident angle changes $\Delta\alpha$ within the memory effect range, the speckle pattern at a distance $v$ from the scattering medium shifts linearly $\Delta l = v \cdot \Delta\alpha$. In other words, the speckle pattern generated by a point source (i.e. the point spreading function, PSF) shifts $\Delta l = v \cdot \Delta x / u$ if the point source shifts $\Delta x$ transversely and $u$ is the distance from the point source to the scattering medium ($\Delta\alpha \approx \Delta x / u$). The shift-invariance of PSFs, or spatial correlation of speckle patterns, allows phase retrieval algorithm to recover the object through a scattering medium without medium characterization [27-30]. Spatial correlation of PSFs or PSF shift-invariance is the key point in these demonstrations, but the wavelength-dependent response of a scattering medium produces de-correlated PSFs with respect to wavelength, and largely degrades the image reconstruction quality. Therefore, monochromatic pseudo-incoherent light source (i.e. dynamically diffused laser light) was preferred in these experiments. With only one PSF measurement, the object, $O$, behind the scattering medium can be recovered from its speckle pattern intensity, $I$, by deconvolution method [31,32]. In this case, a broadband image of an object can be reconstructed from its speckle pattern, in which the scattering medium plays the role of an imaging lens. Multiple shots with multiple optical filters or single shot with integrated multiple filter as in colour cameras could be used for spectral imaging as conventional technologies. Here, we do single-shot multispectral imaging with just a mono-chromatic camera and no optical filter by utilizing the spectral decorrelation effect in scattering media.

Within the memory effect region of a scattering medium, the PSF is linear shift-invariant. Therefore, the captured speckle pattern $I_\lambda$ after the scattering medium is the convolution (denoted as $*$) of an object with the optical system. We can express it as: $I_\lambda = O_\lambda * PSF_\lambda$. The object $O_\lambda$ can be recovered from its speckle pattern $I_\lambda$ by deconvolution: $O_\lambda = deconv(I_\lambda, PSF_\lambda)$, if we know the PSF at wavelength λ of the optical system. The deconvolution process can ideally be expressed as

$$deconv(I_\lambda, PSF_\lambda) = FFT^{-1}\left(\frac{FFT(I_\lambda)FFT(PSF_\lambda)^c}{|FFT(PSF_\lambda)|^2}\right) \quad (1)$$

where $(.)^c$ is the complex conjugate, $FFT(.)$ and $FFT^{-1}(.)$ are the Fourier transform and its inverse, respectively. The convolution in spatial domain becomes the multiplication in Fourier domain, $FFT(O_\lambda * PSF_\lambda) = FFT(O_\lambda)FFT(PSF_\lambda)$ which makes the deconvolution in Eq. (1) possible. As a result of spectral decorrelation in strongly scattering media, the speckle patterns are wavelength dependent; and spectrally separated light sources produce uncorrelated speckle patterns. Using this spectrum dependent behavior of strongly scattering media, spectrometers with high spectral resolution have been demonstrated [33-35]. This spectral decorrelation can be mathematically expressed as follows.

$$PSF_{\lambda 1} \star PSF_{\lambda 2} = \begin{cases} 0 & if \quad \lambda 1 \neq \lambda 2 \\ \delta & if \quad \lambda 1 = \lambda 2 \end{cases} \quad (2)$$

where $\star$ is the correlation operator, and $\delta$ is the spatial impulse function. And the speckle pattern of the multispectral object is a composite response of the scattering medium with all the wavelengths in the object's spectral bandwidth [Fig. 1(a)] as: $I = \sum_\lambda I_\lambda = \sum_\lambda (O_\lambda * PSF_\lambda)$. The following relationship can be deduced: $FFT(I)FFT(PSF_\lambda)^c = FFT(I_\lambda)FFT(PSF_\lambda)^c$ because of the relation: $FFT(A \star B) = FFT(A)FFT(B)^c$. Therefore, each spectral band of the object can be reconstructed from a single monochromatic image $I$ as follow.

$$O_\lambda \approx deconv(I, PSF_\lambda) \quad (3)$$

In essence, each $PSF_\lambda$ or spectral PSF [Fig. 1(b)] not only is for image reconstruction by deconvolution but also plays the role of a spectral filter. More importantly, the single speckle image $I$ of object essentially contains both the spatial and spectral information which are multiplexed orthogonally via multispectral PSFs [Fig. 1(a)]. The deconvolution algorithm will de-multiplex and recover the hidden spectral images. Figure 1(c) presents 9 individual spectral images recovered from 9 corresponding spectral PSFs [Fig. 1(b)] with very high fidelity. A full-spectrum image of the object is presented in Fig. 1(d) by superimposing these multispectral images. It is worth to note that the single wavelength λ presented in this principle can be extended to be a spectral band, and different values of λ present different disjointed bands.

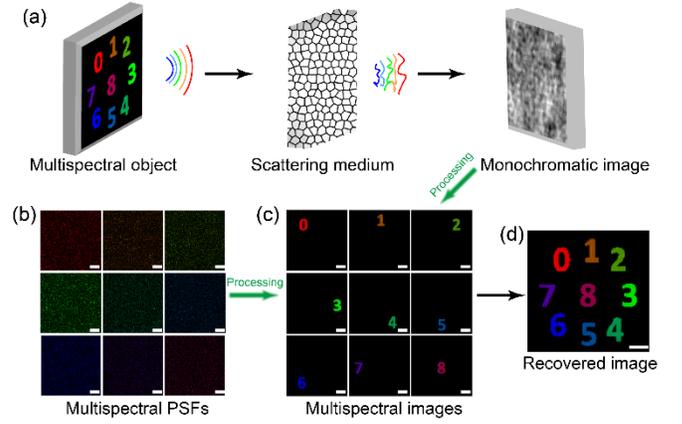

**Fig. 1.** Schematic of our multispectral imaging technique with a scattering medium and a monochromatic camera: simulation results. (a) Light from a multispectral object propagating through a strongly scattering medium generates a speckle pattern on a monochromatic camera. (b) The speckle patterns produce by a central point object illuminated with different spectral bands are recorded as spectral PSFs. (c) Reconstructed spectral images from the monochromatic speckle image using corresponding spectral PSFs. (d) A full-spectrum image of the object is created by superimposing of individual spectral images. Scale bars: 20 pixels.

For demonstration, we generate various 2D multispectral objects [Fig. 2(a)] with three spectral bands corresponding to three primary colours (RGB) of the projector display technology (details about the experiments are presented in Section 1, Supplement 1). Three letters NTU in RGB colour respectively are chosen for the first experiment and their raw speckle pattern captured by the monochromatic camera is presented in Fig. 2(b). The size of this 3-letter object is 1.5 by 0.4 mm, which is well inside the field of view (3.2 mm) for our optical imaging system defined by the memory effect of the scattering medium (Fig. S1, Supplement 1). We display 4 different colours (RGB and white) on the central pixel of projector while shutting down all the other pixels to generate the effective multispectral point sources. Then the camera takes their speckle images as multispectral PSFs for the imaging system. Figure 2(c) (upper) shows the recovered multispectral images by deconvolving the speckle pattern in Fig. 2(b) with corresponding spectral PSFs. Each spectral PSF successfully reconstructs the respective letter. The broadest-spectral PSF (white PSF) reconstructs all the three letters NTU. This is expected because in projector display technology, the white pattern and the white PSF are composition of three individual RGB patterns and RGB PSFs, respectively. We notice that a dim letter U appears in the green spectral image, and a dim letter T appears in the blue spectral image. This is confirmed in the intensity of the central row [Fig. 2(c) - lower] and in Fig. S2(a)-(c), Supplement 1. We superimpose the 3 RGB spectral images and achieve the "full colour" image in Fig. 2(d), successfully recovering the original multispectral object NTU with magnification of $v/u \approx 0.42$.

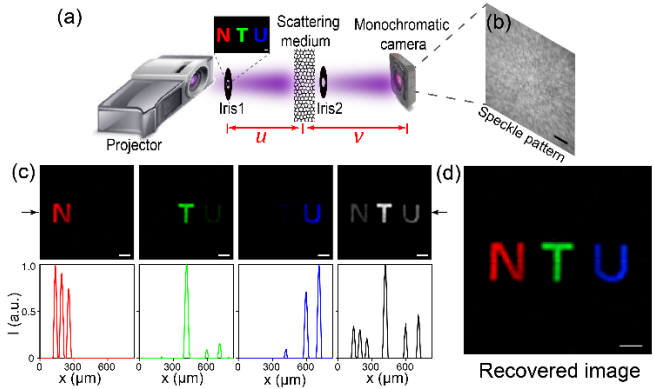

**Fig. 2.** Experimental setup and the reconstructed multispectral images. (a) A common projector without a magnification lens generates 2D multispectral objects on the plane of iris-1 (the object plane), which blocks any stray light from projectors. Light path is scrambled by a scattering medium (optical diffuser) then goes to a monochromatic camera after passing through the second iris, which plays the role of aperture in our imaging system. (b) The raw speckle pattern captured by the monochromatic camera. (c) Reconstructed multispectral images (upper half) and the intensity of their central rows (lower half), respectively. Two arrows mark the central row of the images. (d) A composite multispectral image is constructed by superimposing three individual RGB spectral images in (c). Scale bars: 1000 μm in (b) and 100 μm in others.

The "cross-talk" effect between green and blue channels in fact comes from a relatively high cross-correlation coefficient of 0.1941 between these blue and green spectral PSFs [Fig. 3(a)]. On the other hand, the small cross-correlation coefficients of the red spectral PSF with the green and blue ones (0.0691 and 0.0255, respectively) only result in the background noise of the spectral images. From our mathematical principle in Eq. (2), spectral PSFs are playing the role of spectral filters in which the cross-correlation coefficient between two spectral PSFs presents the "transmission coefficient" of a spectral band via the other spectral band filter. This principle explains the observations in the recovered white spectral image. The brightest letter T corresponds to the highest cross-correlation coefficient between the green and white spectral PSF (0.8660). And almost similar cross-correlation coefficients of the white spectral PSF with the red and blue ones (0.4536 and 0.4374, respectively) result in similar intensities for reconstructed letter N and U.

To understand the nature of cross-talk in our PSFs, we analyze the spectra of different bands produced by our projector [Fig. 3(b)]. The white spectrum is indeed the sum of the three RGB spectral bands. However, the RGB bands are not spectrally disjointed. The blue band has a significant spectral overlap with the green band, justifying their high PSFs' correlation. There is no spectral overlap of blue and red spectral PSFs, making their cross-correlation coefficient negligible. The green spectrum contributes the most intensity to the white spectrum, and therefore the green PSF has the highest cross-correlation coefficient with the white PSF. The appearance of a dim letter U and T in the green and blue recovered spectral images, respectively, illustrates the ability of our technique to spatially and spectrally resolve a weak signal from a multispectral object. To demonstrate the spectral filtering role of the PSFs further, we used narrower RGB band-pass filters in front of the camera when measuring PSFs to avoid any spectral overlap [Fig. 3(b)]. These narrow-spectral PSFs show negligible cross-correlation coefficients [Fig. 3(c)] as expected from their disjointed spectra and spectral decorrelation effect in our strongly scattering medium. Figure 4(a) shows the reconstructed RGB spectral images, which are de-convolved from the same speckle image in Fig. 2(b) with the new narrow-spectral PSFs. Three clean letters

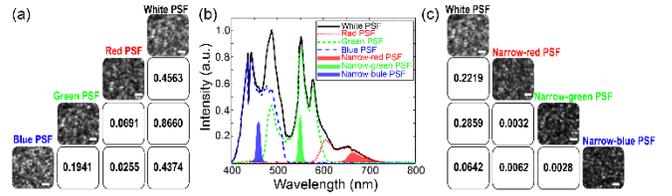

**Fig. 3.** The relationship between cross-correlation and spectral overlap of spectral PSFs. (a) The white, RGB spectral PSFs and their cross-correlation coefficients. (b) Spectra of the four spectral PSFs (RGB and white) together with three narrow-band PSFs at RGB colour, respectively. Three narrow band-pass filters are added in front of the projector consequently then we measure the transmitted spectra. (c) The white PSF and RGB narrow-spectral PSFs together with their cross-correlation coefficients. Scale bars: 100 μm.

NTU appear exactly at their corresponding spectral images without any cross-talk as illustrated in their intensity at the central row [Fig. 4(b)] as well as in their heat map version [Fig. S2(d)-(f), Supplement 1]. More importantly, this demonstration shows that the multispectral images with any bands are embedded in the raw speckle image [Fig. 2(b)], we just need multispectral PSFs to retrieve desired multispectral images. Each optical configuration (an optical diffuser and a monochromatic camera) has unique and unchanged spectral PSFs, which only need to be measured once for all samples and analysis. The spectral response of the imaging system is essentially defined by the camera and absorption coefficient of the scattering medium. Using our current camera, we can do single-shot multispectral imaging of any spectral band within a broad range from UV (250 nm) to near IR (1100 nm) just by adding a glass diffuser. These advantages differentiate our technique from any existing spectral imaging technologies.

Our approach provides an additional dimension of spectrum to a 2D imager, which has an upper bound limitation on information that can be captured and then segregated. Within this limit, we have to trade-off between spatial and spectral information, the spectral range and spectral resolution. However, the scattering media give us the choices for flexible implementation to utilize the full capacity of the 2D imager. We can limit the spectral information to a single narrow line and get maximum spatial information, i.e. maximum number of spatial pixels and dynamic range. The number of active points and the number of de-correlated spectral bands in the object determine the speckle contrast and therefore the recovery quality of our technique. As a result, we can easily retrieve large amount of spatial information with high spectral resolution for sparse spectral lines in a broad spectral range.

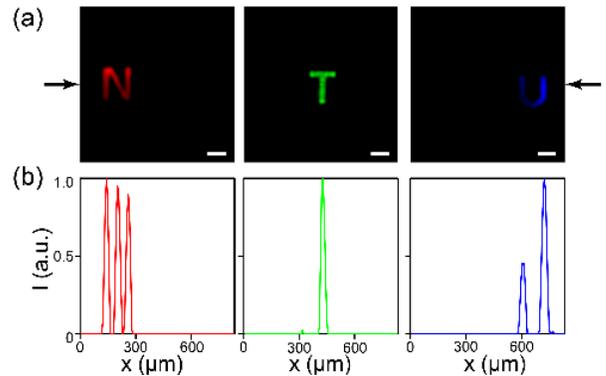

**Fig. 4.** Narrow-band multispectral imaging analysis. (a) Reconstructed multispectral images from the raw speckle image in Fig. 2(b) with corresponding narrow-spectral PSFs. Scale bars: 100 μm. Two arrows mark the central row of the images. (b) Intensity across the central row of the corresponding images in (a).

On the other extreme, we can limit our spatial information to a single point object, and then retrieve highest spectral resolution of the point source. We numerically simulate and present this spectrometer application of our optical system in in Fig. S4, Supplement 1. The reconstructed spectra of the white light emitting diode (LED) at different samplings of $\Delta\lambda$ = 2 nm and 10 nm illustrate a potential spectroscopy application in our flexible multispectral imaging device. It is important to note that our multispectral imaging technique is insensitive to optical alignment because of no moving part, no focusing optics. Due to memory effect, a small shift in the optical setup only causes a small shift of reconstructed multispectral images and we do not need to retake the multi-spectral PSFs. Our prototype with a diffuser, an iris mounted to the camera through a lens tube, has been working very well for more than a month with one time measurement of multispectral PSFs at the beginning. The level of spectral decorrelation and spatial correlation effect together with the resolution (pixel size and pixel number), dynamic ranges and noise level of the monochromatic camera define the performance of our technique (Section 5-8, Supplement 1). More investigations of new scattering media with stronger spectral decorrelation and spatial correlation will improve this cost-effective technology significantly and create great impacts in many important applications.

We demonstrate a multispectral imaging technique by just adding a strongly scattering medium in front of a monochromatic camera. A single-shot speckle pattern essentially contains spatial and spectral information of the object. The multispectral images with any spectral bands can be retrieved with the corresponding spectral PSFs. The respective spectral PSFs are fixed and only need to be recorded once for all time use. The deconvolution algorithm utilizes the shift invariance (i.e. spatial correlation) of the PSFs for image reconstruction, while the orthogonality between the spectral PSFs (i.e. spectral decorrelation) makes them play the role of spectral filters. Our demonstrations present the simplest technique for spectroscopy and multispectral imaging that allows flexible trade-off between spatial and spectral information, as well as spectral range and spectral resolution to be captured and retrieved in a single grey-scale image.

**Funding.** NTU start-up grant; Singapore MOE-AcRF Tier-1 grant (RG70/15); Singapore Ministry of Health's National Medical Research Council under its <CBRG-NIG (NMRC/BNIG/2039/2015)>.

**Acknowledgment**. We would like to thank Yu Tian, Vinh Tran and Dr. Balasubramanian Padmanabhan for fruitful discussions and useful feedback. The research is supported by the NTU start-up grant, Singapore MOE-AcRF Tier-1 grant (RG70/15) and the Singapore Ministry of Health's National Medical Research Council under its <CBRG-NIG (NMRC/BNIG/2039/2015)>.

See Supplement 1 for supporting content.

# Single-shot multispectral imaging with a monochromatic camera: supplementary material

**SUJIT KUMAR SAHOO,**[1,2,*] **DONGLIANG TANG,**[1] **AND CUONG DANG**[2,*]

[1]*Centre for OptoElectronics and Biophotonics (OPTIMUS), School of Electrical and Electronic Engineering, The Photonic Institute (TPI), Nanyang Technological University Singapore, 50 Nanyang Avenue, 639798, Singapore*
[2] *Department of Statistics and Applied Probability, National University of Singapore, Singapore*
*\*Corresponding author: sujit@pmail.ntu.edu.sg, hcdang@ntu.edu.sg*

This document provides supplementary information to "Single-shot multispectral imaging with a monochromatic camera,". It contains detailed information regarding the optical setup and the post-processing for the data, measurements for the field of view, demonstrations for mixed color components and a demonstration for the spectroscopy technique with scattering media. In addition, we analyze how the possible factors, such as the image dimension, spectral bands, accuracy of point spreading function and noises affect the reconstructions. © 2017 Optical Society of America

## 1. Optical Setup and Data Processing

A common projector (Acer X113PH) without the second projection lens is utilized to generate 2D multispectral objects at the plane of iris-1 [Fig. 2(a)]. The small lens inside the projector makes the magnification of $M_1 \approx 1.82$ from its DMD chip to the object plane. The diffuser (Edmund, 120 Grit Ground Glass Diffuser) combined with the iris-2 are positioned at a distance $u \approx 210mm$ from the object plane. The diameter of the iris-2 is about 2mm to obtain an appropriate speckle intensity and signal-to-noise ratio. The light coming from the 2D object passes through the diffuser and generates the speckle pattern on a high-dynamic monochromatic CMOS (Andor Neo 5.5, 2560×2160, pixel size 6.5um), which is placed at a distance $v \approx 87.5mm$ from the diffuser. Therefore, the magnification of our imaging setup is $M_2 = v/u \approx 0.42$ and the total magnification from the DMD chip to CMOS sensor is $M_{total} = M_1 \cdot M_2$. We apply the Wiener deconvolution algorithm, which takes about 0.5 second for each spectral image to be reconstructed, using MATLAB on a normal PC (Intel Core i7, 16 GB memory). Spectral PSFs are constant for an optical system and their FFT can be pre-calculated and stored to reduce the reconstruction process by 30% (Visualization 1). The central area with 2048*2048 pixels on the CMOS sensor is chosen for our experiments. All the speckle images and PSFs are grey images as they are captured by a monochromatic camera. We set colours into these multispectral PSFs and multispectral images for our illustration purpose only.

For each speckle patterns and PSFs, we divide them with their low frequency envelop to remove the hallow effect and sharpen the speckles. The superimposed images are created from spectral images with our defined colours for the illustration purpose as well. All the superimposed images (colour images) are composite of 3 reconstructed red, green and blue (RGB) spectral images without any colour processing. The spectra of white, and RGB light from the projector are measured by Avantes spectrometer (AvaSpec-2048).

## 2. Measurement for the field of view (FOV)

Any object located in the memory effect region of the scattering medium generates the shift-invariant speckle pattern on the camera. This memory effect region is the incident angle region ($\phi$), defining the field of view (FOV) in our experiment: $L = \phi u$, where u is the distance from object plane (iris-1's plane) to the scattering medium. The "ON" pixels on the digital micro-mirror device (DMD) of projector define the 2D object on the object plane with magnification of $M = 1.82$. We characterize the FOV by measuring the cross-correlation coefficient between the PSF at the centre and PSFs at different positions (x). Figure S1 presents the PSF cross-correlation coefficient as a function of point's position on the object plane. If we define the FOV is the region with PSF cross-correlation coefficient of greater than 0.5, the FOV will be inside the circle with diameter of about 3.2 mm. We also can estimate the effective thickness of scattering medium (L) by fitting the experiment data with the theoretical model for the memory effect as follows [1]:

$$C(\theta, L) = [k_0 \theta L / \sinh(k_0 \theta L)]^2 \quad \textbf{(S1)}$$

where $C$ is the correlation coefficient, $k_0$ is the wave vector ( $k_0 = 2\pi / \lambda$ ), and $\theta = x/u$ is the incident angle. If we use light at wavelength (λ) of peak intensity which has the main contribution in creating a PSF to calculate $k_0$, we can estimate roughly the effective thickness of our scattering medium of about: 18, 22, 18 and 16 μm for white and RGB band, respectively.

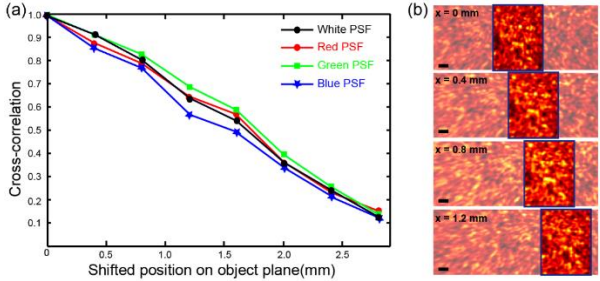

**Fig. S1.** Memory effect in our strongly scattering medium. (a) Cross-correlation coefficient between the PSF at the centre (x = 0) and PSFs at other positions. The experiment is repeated for four spectral bands: RGB and white. (b) Selected region of white PSFs at different positions. Highlighted regions show the shift-invariant speckle patterns when the white colour pixel at the positions of x = 0, 0.4, 0.8, 1.2 mm, respectively. Scale bars: 100μm.

## 3. Effect of spectral overlap on the reconstructed multispectral images

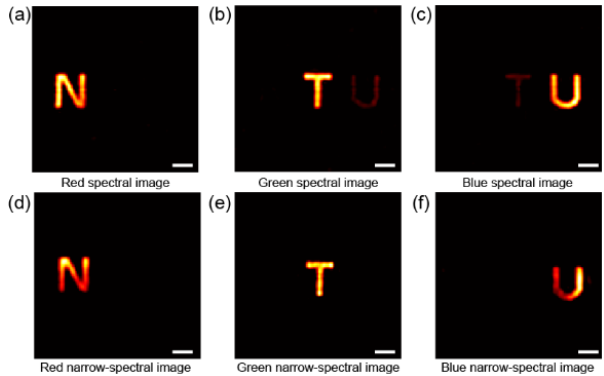

**Fig. S2.** Effects of spectral overlap in PSFs, presented in heat-map images. (a-c) The RGB spectral images, respectively, i.e. the heat-map version of Fig. 2(d) for RGB images only. The cross-talk effect is observed between green and blue spectral channels due to their spectral overlap. (d-f) The RGB narrow-spectral band images, i.e. the heat-map version of Fig. 3(c), respectively. Non-overlap in spectra of these RGB narrow-spectral PSFs [Fig. 3(d)] makes separated N, T, U letters in three RGB spectral images. However, the weak intensities of a single projector's pixel with narrow band filters in green and especially blue region make pure measurements (very low SNR) of narrow-spectral PSFs. This low SNR reduces the quality of reconstructed spectral images in narrow green (e) and especially narrow blue (f) band. Scale bars: 100 μm.

## 4. Demonstrations for mixed color components.

We further demonstrate our technique with more complex spectral information in Fig. S3. Three letters CDG [Fig. S3(a)] presented in cyan (green and blue), magenta (red and blue), and yellow (red and green), respectively or a complex multispectral object with gradient and blended colours [Fig. S3(b)] are utilized to highlight the capabilities of our technique. Four spectral PSFs as used for recovering in Fig. 2(c) are employed to de-convolve the raw speckle images in Fig. S3(c)-(d). The reconstructed multispectral images in Fig. S3(e) presents successful recovery of the pairs of letters that shares the same spectral element in Fig. S3(a) by the respective RGB spectral PSFs. Deconvolution with the white spectral PSF results in three letter CDG as expected with strong intensity at letters C and G because of the strong green intensity in white spectrum [Fig. 3(b)]. Figure S3(f) shows the recovered multispectral images of object in Fig. S3(b) with gradient intensities. The final composited "rainbow" colour images are displayed in Fig. S3(g)-(h), resembling the original complex multispectral object in Fig. S3(a)-(b), respectively. The colour quality is distorted by the wavelength dependence of camera sensitivity, which is highest at green and lowest at blue range. Calibrating with sensitivities of the camera and human eyes in a colour processing technique as normal colour cameras do will enhance the colour appearance of superimposed images to be similar to the original object.

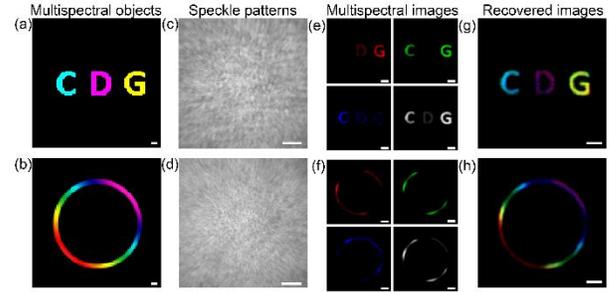

**Fig. S3.** Reconstruction of mixed and gradient multispectral objects. (a) Multispectral 'CDG' letters with corresponding cyan, magenta and yellow (CMY) colour are generated as an object. In display technology, CMY colours are generated by combination of three primary RGB colours: cyan is a mixture of green and blue, magenta is a mixture of red and blue, yellow is a mixture of red and green. (b) Another object of a circular ring with gradient rainbow colours. (c), (d) The raw speckle patterns captured by the monochromatic camera for the object in (a) and (b), respectively. (e), (f) Reconstructed spectral images from the raw speckle images in (c) and (d), respectively. We use the corresponding spectral PSFs in Figure 2(c) for the deconvolution processes. (g), (h) The composite spectral image of the object is created by super positioning three RGB spectral images from (e) and (f), respectively. Scale bars: 1000 μm in (c) and (d); 100 μm in others.

## 4. A spectroscopy technique with strongly scattering media

We now demonstrate the ability of our technique to resolve spectral information of a point source. Here, we pass a broadband light point source through the scattering medium, and measured speckle intensities using a monochromatic camera. Then, we construct the spectrum of the broadband point source by calculating the cross-correlation coefficient between the the multispectral PSFs with the measured broadband speckle image. One can use a broadband light source with a monochromator to measure various spectral PSFs at different wavelengths. This one-time "calibration" step could be done at a manufacture. To demonstrate the proof of concept, we numerically simulate various spectral PSFs as independent speckle patterns. The number of spectral PSFs is the product of spectral sampling rate (sample/nm) and the spectral bandwidth. Then, the speckle for a broadband point source is generated by weighted combination of these independent spectral PSFs, where the weights equal to the intensities of the respective spectral lines. We measure the spectrum of a commercial white LED and use it as the broadband for our demonstration. The recovered spectra and the original spectrum is plotted together in Fig. S4. The results show successful recovery of actual spectrum. We simulate here with different spectral sampling rates of Δλ = 2 nm and 10 nm, in which the PSFs for λ and λ+Δλ are independent, i.e. de-correlated. In reality, how much they are de-correlated and how well our camera detects their differences will define the spectral resolution of our approach.

Special design of scattering media can achieve the spectrometer resolution of sub-nanometer [2-5]. For normal optical diffuser or biology tissue, Ori Katz et al. used a tune-able laser to characterize the spectral decorrelation bandwidth of 5 - 10 nm for 20μm-thick sample of $TiO_2$ particles of diameter <250nm or 500μm-thick chicken breast in their optical setup [6]. For any specific scattering medium, increasing effective thickness will reduce the spectral decorrelation bandwidth, therefore increase the spectral resolution in our spectrometer applications. Nevertheless, the 5~10 nm resolution of the normal diffuser is already useful for many purposes. For example, one can make a simple spectrometer with scattering medium and a smartphone to characterize our LED spectrum.

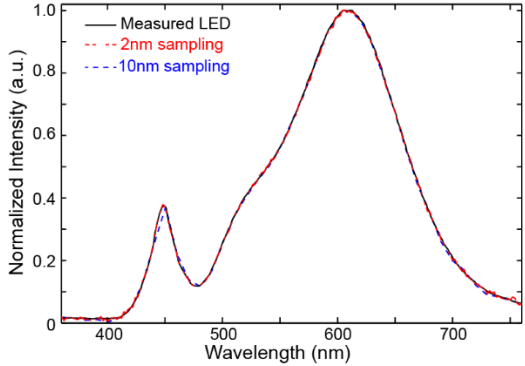

**Fig. S4.** Demonstration of spectrometer application utilizing spectral decorrelation of PSFs through a scattering medium. Spectrum reconstruction with various separation spectral lines (2 nm and 10 nm, respectively).

## 5. Deconvolution in presence of noise

The notion of deconvolution has been introduced in the main text and the previous section to have simple illustration of the principle. However, the deconvolution is not just an ideal inversion of the PSF. In the practice, we have to operate in the finite precision and within a fixed dynamic range of the system. In order to handle all these practical constrains, the observed image is mathematically expressed as

$$I = I_{ideal} + V \quad \text{(S2)}$$

where $I_{ideal} = \sum_\lambda O_\lambda * PSF_\lambda$, and $V$ is modeled as an additive white Gaussian noise of variance $\sigma^2$. Such kind of model can lead us to a very close approximation to the PSF inversion. There exist numerous deconvolution techniques to solve the problem model [7-10]. Weiner deconvolution is one of the fastest technique, which we use throughout our simulations and experiments [7]. The formula for the deconvolution is the following.

$$O_\lambda = weiner(I, PSF_\lambda) = FFT^{-1}\left(\frac{FFT(I)FFT(PSF_\lambda)^C}{|FFT(PSF_\lambda)|^2 + C\sigma^2}\right) \quad \text{(S3)}$$

where $C>1$ is a constant to keep the approximate solution clear from the measurement deformation artifacts. The Weiner deconvolution formula is also very intuitive and it has close resemblance with the ideal deconvolution (Eq. S3). In practice the noise variance of the imaging system/environment can be measured and the value of $C$ can be determined.

There are many existing iterative deconvolution algorithms such as Richardson–Lucy that doesn't need the noise variance parameter [10-12]. However, the noise estimation and reduction is automatically taken care within the iteration loops. As a result, they produced heavily smooth images with other artifacts. The main drawbacks of these iterative algorithms are the execution time that grows in polynomial-scale with total pixels ($N$) and needs many iteration, whereas Weiner algorithm's execution time is in logarithmic-scale of $N$ without any iterations. The study of various deconvolution methods is beyond the scope of this work, but it might be worth to test our principle with various deconvolution techniques.

## 6. Effect of imager's dimensionality

All the measurement and finite precision noise are always modeled as independent additive noise. However this assumption is only valid in infinite dimension and infinite precision. As a result, all the convolution algorithms suffer from reconstruction artifacts depending on the amount the correlation between the noise and the signal. In principle, the correlation is a decreasing function [13] of total pixels ($N$). To demonstrate that we plot the cross-correlation's value of two independent random signals with respect to dimension, taking an average over 1000 trials.

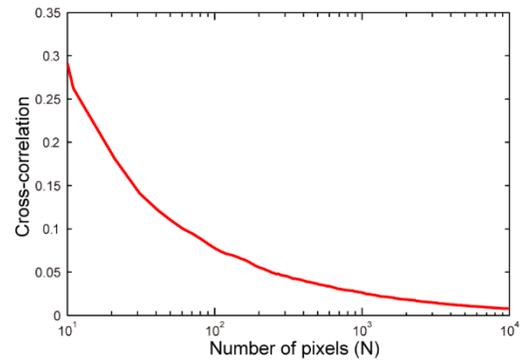

**Fig. S5.** Cross-correlation coefficient of two independent random images at various number of pixel.

To visually illustrate the effect of dimensionality on the reconstruction artifacts, we have reconstructed the object by cutting out various size of the measured speckle image and the measured PSFs. We simulate independent speckle patterns for RGB spectral PSFs. Each color component is convolved with the respective PSFs and summed up to form the measured gray scale image. Each of the color component was extracted by deconvolving the speckle image with the respective PSF, and the final image is displayed by superimposing of each reconstructed spectral image in the respective colour channel. The following results are displayed for the various sizes of deconvolution kernels. We can see that the increasing dimension improves quality of the reconstruction image.

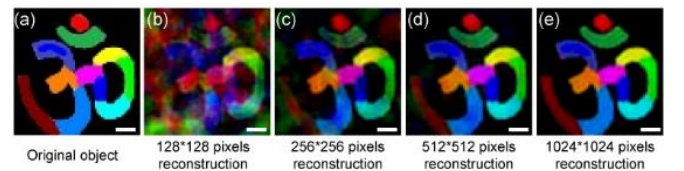

**Fig. S6.** Effect of dimensionality in our multispectral imaging technique. (a) An original object of 64x64 pixel. (b-e) The reconstructed multispectral images from the cropped raw speckle image and cropped PSFs at different sizes: 128x128, 256x256, 512x512, 1024x1024, respectively. Scale bars: 10 pixels.

## 7. Effect of total spectral band numbers

The same notion on increasing correlation can be extended to the principle of multispectral imaging introduced in this letter. The multispectral imaging relies on the spectral decorrelation property of the speckle patterns. This allows us to segregate the superimpose image of each spectral band. The super position can be described as the following.

$$I = O_{\lambda 1} * PSF_{\lambda 1} + (\sum_{\lambda \neq \lambda 1} O_\lambda * PSF_\lambda) + V$$
$$= O_{\lambda 1} * PSF_{\lambda 1} + (V_{spectra} + V) \quad \text{(S4)}$$

We assumed that the PSF of the recovering bands are uncorrelated to each other (i.e. $PSF_{\lambda 1} \star PSF_\lambda \approx 0$, if $\lambda_1 \neq \lambda$). Therefore, the speckle intensity component $V_{spectra} = \sum_{\lambda \neq \lambda 1} O_\lambda * PSF_\lambda$ is treated as secondary noise. This is very much viable in a larger finite dimension. However, as the number of band increases the component $V_{spectra}$ increases, which will cause a linear rise in the correlation. As a result, we will have some reconstruction artifacts, which can be suppressed with over-smoothing of the reconstructed images [10].

### 8. Accuracy of Point Spreading Function

In the previous section, we have discussed the importance of noise, correlation, dimensionality and number of spectral bands. We have safely assumed the noise only appears during the measurement of the speckle image I. This assumption is quite practical, because the PSF is a one-time measurement. Therefore, more resources such as high-power wavelength tune-able light sources can be devoted to have very accurate spectral PSF measurement at the manufacture. However, our current setup with small projector, the light intensity from a single pixel is very low and our spectral PSFs present some noise. The signal to noise ratio (SNR) is even lower when we use narrow band-pass filters to acquire narrow-spectral PSFs. As a result, we can notice the reconstruction artifacts for some of the bands in our reported results [Fig. S2(f)]. The reconstruction with a noisy PSF can be expressed as follows.

$$O_\lambda = FFT^{-1}\left(\frac{FFT(I)FFT(PSF_\lambda + V_\lambda)^C}{|FFT(PSF_\lambda + V_\lambda)|^2 + C\sigma^2}\right) \quad \text{(S5)}$$

Taking the assumption of the uncorrelated PSF noise $V_\lambda$, we can further simplify the expression as follows.

$$O_\lambda \approx FFT^{-1}\left(\frac{FFT(I)FFT(PSF_\lambda)^C}{|FFT(PSF_\lambda + V_\lambda)|^2 + |V_\lambda|^2 + C\sigma^2}\right) \quad \text{(S6)}$$
$$+ [smoothed\ noise]$$

This deduces that the resulting image would be slightly over smoothed due to $|V_\lambda|^2$, and it would have some smoothed noise due to finite dimension correlation artifact. A similar imaging experiment as the above section is repeated again for various noise level of the PSF used for reconstruction. The image is generated using the ideal PSFs for convolution, but we have reconstructed the multispectral images with the noisy green PSF. Images are reconstructed for various SNR levels, and the results are displayed in Fig. S7 below. We can observed the degraded reconstruction with the increase in PSF noise.

reconstructed multispectral images when the green PSF is corrupted by noise at different SNR level: 5dB, 10dB, 15dB, and 20dB, respectively. Scale bars: 10 pixels.

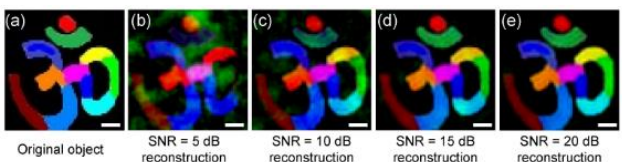

**Fig. S7.** Effect of noise in PSF measurement on reconstructed multispectral images. (a) An original object of 64x64 pixel. (b-e) The